\documentclass[aps,prd,twocolumn,nofootinbib,superscriptaddress]{revtex4-2}

\usepackage{amsmath,amssymb}
\usepackage{graphicx}
\usepackage{xcolor}
\usepackage{hyperref}
\usepackage{float}
\usepackage{bm}

\hypersetup{
    colorlinks=true,
    linkcolor=blue,
    citecolor=blue,
    urlcolor=blue
}

\DeclareMathOperator{\Mwh}{M_{\mathrm{WH}}}

\begin{document}

\title{Gravitational Waves from a Black Hole Falling Radially into a Thin-Shell Traversable Wormhole}

\author{Mohammad Nosherwan Malik}
\email{mohammad.nosherwan.malik@vanderbilt.edu}
\affiliation{Department of Physics and Astronomy, Vanderbilt University, Nashville, TN 37235, USA}

\author{James B. Dent}
\email{jbdent@shsu.edu}
\affiliation{Department of Physics, Sam Houston State University, Huntsville, TX 77341, USA}
 
\author{William E. Gabella}
\email{bill.gabella@gmail.com}
\affiliation{Department of Physics and Astronomy, Vanderbilt University, Nashville, TN 37235, USA}

\author{Thomas W. Kephart}
\email{tom.kephart@gmail.com}
\affiliation{Department of Physics and Astronomy, Vanderbilt University, Nashville, TN 37235, USA}

\date{\today}

\begin{abstract}
We compute the gravitational-wave signal generated by the radial infall of a stellar-mass black hole into a thin-shell Schwarzschild traversable wormhole. Modeling the black hole as a test particle, we derive analytic expressions for the emitted waveform, including contributions from the mass quadrupole and higher-order multipoles. The resulting signal exhibits a characteristic pulse--gap structure associated with repeated throat crossings. We further compute the amplitude spectral density and compare it with representative ground-based detector sensitivities, finding that such signals could lie within the sensitivity range for optimally oriented sources at distances of order $\sim 500\,\mathrm{Mpc}$. These results provide a potential observational signature of traversable wormholes in gravitational-wave data.
\end{abstract}

\maketitle

\section{Introduction}

The first direct detection of gravitational waves (GWs) from a binary black hole merger by the LIGO Scientific Collaboration marked a historic milestone in astrophysics, confirming a central prediction of general relativity and opening a new observational window onto the Universe~\cite{LIGOScientific:2016aoc}. Since that initial discovery, the LIGO–Virgo–KAGRA collaboration has reported dozens of additional compact-binary mergers across multiple observing runs, summarized in the gravitational-wave transient catalogs GWTC-1 through GWTC-3~\cite{LIGOScientific:2018mvr,LIGOScientific:2020ibl,LIGOScientific:2021djp}. These observations have transformed gravitational-wave astronomy into a precision discipline, enabling detailed tests of strong-field gravity and compact-object population studies.

In parallel, pulsar timing arrays, including NANOGrav, have recently reported evidence for a stochastic nanohertz gravitational-wave background~\cite{NANOGrav:2023gor}, opening a complementary low-frequency window onto supermassive black hole binaries and possible cosmological sources.

Even prior to direct detection, it was recognized that gravitational waves could serve not merely as confirmations of general relativity but as precision probes of spacetime in extreme regimes~\cite{Thorne:1995xs}. Theoretical developments have further connected gravitational-wave physics with deep questions about the quantum structure of spacetime and black hole interiors~\cite{Maldacena:2013xja,Deng:2016vzb,Deng:2017uwc,Flores:2024lng,Dent:2025bwo,Ning:2026nfs}. Beyond astrophysical binaries, gravitational waves have been proposed as probes of a wide variety of new physics including dark matter~\cite{Miller:2025yyx,Chen:2026mef,Bertone:2024rxe}, first-order phase transitions in the early Universe~\cite{Weir:2017wfa,Mazumdar:2018dfl,Hindmarsh:2020hop,Athron:2023xlk,Caprini:2025trt}, primordial black holes~\cite{Carr:1974nx,Bird:2016dcv,Nakama:2016gzw,Sasaki:2018dmp,Green:2020jor,Papanikolaou:2020qtd,Carr:2021bzv,Carr:2026hot}, and horizonless or exotic compact objects~\cite{Pani:2009ss,Cardoso:2016oxy,Cardoso:2019rvt}. Together, these advances establish gravitational waves as a powerful laboratory for exploring both strong-field gravity and physics beyond the Standard Model. 

The concept of traversable wormholes was introduced in a modern general relativity framework by Morris and Thorne as hypothetical structures enabling rapid interstellar travel~\cite{Morris:1988cz} (for other early wormhole investigations, see also~\cite{Ellis:1973yv,Bronnikov:1973fh,Sato:1981bf,Maeda:1981gw}). Since then, wormholes have been studied extensively~\cite{Bambi:2021qfo}, and gravitational waves have been proposed as a possible observational probe of their existence~\cite{Bao:2022iaz}. Test particle motion in thin-shell wormhole geometries has been analyzed to understand matter behavior near such exotic structures, revealing that stable orbits can exist under certain conditions and may provide observational signatures~\cite{Diemer:2013hgn}. Gravitational waves from the inspiral of a stellar-mass black hole into a stable, non-spinning traversable wormhole have also been modeled~\cite{Dent:2020nfa}, showing that the waveform may exhibit a characteristic \textit{chirp/anti-chirp} pattern and/or a \textit{burst} as the black hole emerges—i.e., outspirals—into our region of the Universe.

In this paper, we investigate the gravitational waves emitted during the radial infall of a stellar-mass black hole into a stable, non-spinning, traversable wormhole. Specifically, we consider the motion of the black hole along a single spatial axis, beginning in our region of the Universe and passing through the wormhole throat into either a second asymptotically flat universe (Universe 2) or a distant region of the same universe. We calculate the resulting gravitational wave strain—both the plus $h_+$ and cross $h_\times$ polarizations—arising from this dynamic system. Our analysis includes contributions from the mass quadrupole, mass octupole, and current quadrupole moments generated by the asymmetric mass-energy distribution during the in-fall. As we show in later sections, the emitted waveform exhibits a distinct \textit{pulse} pattern. The \textit{pulse} arises as the black hole approaches the wormhole throat from our side of the Universe, with increasing frequency and amplitude. Once the black hole passes through the throat into the other region, the waveform reverses its behavior, producing a \textit{burst} as the signal in the exit phase. These distinctive features could potentially serve as observable signatures of traversable wormholes in gravitational wave data.

\section{Wormhole Construction}

For the purposes of this paper, we consider a Morris–Thorne-type Lorentzian wormhole (WH), whose metric functions can be chosen to keep tidal forces arbitrarily small along the infalling black hole’s (BH) trajectory. Such a \textit{traversable wormhole} necessitates matter that violates the weak energy condition (WEC), implying the existence of \textit{exotic matter} with negative energy density~\cite{Morris:1988cz, Morris:1988tu}. We construct this wormhole surgically from two Schwarzschild BH spacetimes and consider a BH of mass $M_{\text{BH}}$ falling into the wormhole of mass $M_{\text{WH}}$, following a straight-line trajectory along the $z$-axis. To ensure that the BH’s passage through the throat does not significantly perturb the wormhole geometry or destabilize the exotic matter shell (as discussed later), we take the mass ratio
\[
\frac{M_{\text{WH}}}{M_{\text{BH}}} \gtrsim \mathcal{O}(10),
\]
 which minimizes backreaction effects and keeps tidal stresses sufficiently small to preserve the throat’s stability during traversal. Following~\cite{Morris:1988tu}, the embedding diagram for this wormhole is shown in Figure~1.

A variety of traversable wormholes that can be constructed via spacetime surgery are discussed in~\cite{Visser:1989am}. In this paper, we focus on a specific example: a thin-shell Schwarzschild wormhole. In this construction, two identical Schwarzschild spacetimes are joined across a timelike hypersurface located at a radius $r = a > 2M$, forming a geodesically complete spacetime without an event horizon or singularity. All of the exotic matter required to sustain the wormhole is confined to an infinitesimally thin layer—modeled as a delta-function source—located at the throat. This localization allows the bulk spacetime to remain Ricci-flat and greatly simplifies the analysis. We adopt Visser’s surgical method~\cite{Visser:1989kh, Visser:1989kg} as it offers both analytic tractability and a physically localized way to handle violations of the energy conditions. The surface stress-energy on the throat is derived using Israel's junction conditions, and its properties can be related to the wormhole’s stability. A detailed analysis of the linearized radial stability of such thin-shell wormholes is given in~\cite{Poisson:1995sv}, where constraints on the wormhole mass, throat radius, and the equation of state of the exotic matter are derived. Extensions of this method to rotating spacetimes have also been explored; notably, Kashargin and Sushkov~\cite{Kashargin:2011fg} constructed a rotating thin-shell wormhole by gluing together two Kerr geometries at a constant Boyer–Lindquist radius.

\begin{figure}[t]
    \centering
    \includegraphics[width=0.7\linewidth]{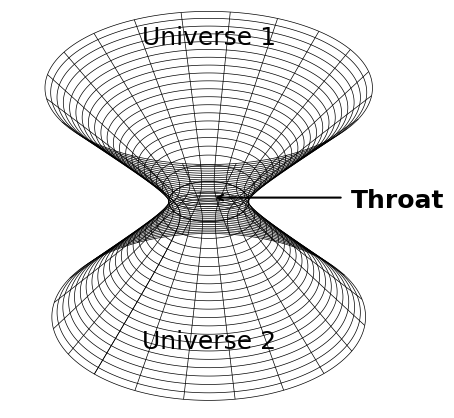}
    \caption{Embedding diagram for the Schwarzschild wormhole. The two asymptotically flat regions—corresponding to Universe 1 (top) and Universe 2 (bottom)—are connected by a smooth throat, which represents the narrowest part of the geometry. For physical consistency, we assume that the throat radius satisfies $r_{\text{throat}} > 2M_{\text{BH}}$, ensuring that the black hole can traverse the wormhole without encountering an event horizon.}
    \label{fig:embedding}
\end{figure}

We start with the Schwarzschild metric in the form (see ~\cite{Hartle:2021pel})

\begin{align}
ds^2 =\ & -\left(1 - \frac{2M}{r}\right) dt^2 
+ \left(1 - \frac{2M}{r}\right)^{-1} dr^2 \nonumber \\
& + r^2 \left( d\theta^2 + \sin^2\theta \, d\varphi^2 \right),
\label{eq:metric}
\end{align}
 and avoid maximally extending the manifold while constructing the wormhole geometry. Instead, we take two copies of the Schwarzschild spacetime and remove the regions defined by \(\Omega_{1,2} \equiv \{r_{1,2} \leq a \mid a > 2M\}\). This leaves us with two geodesically incomplete manifolds, each bounded by timelike hypersurfaces \(\partial \Omega_{1,2} \equiv \{r_{1,2} = a\}\). By identifying these hypersurfaces, i.e., \(\partial \Omega_1 \equiv \partial \Omega_2\), we construct a new spacetime \(\mathcal{M}\) that is geodesically complete and possesses two asymptotically flat regions connected by a wormhole.

At the junction surface \(\partial \Omega\) lies the wormhole throat. The stress energy tensor, taking the form of a delta function, is concentrated at the throat and zero everywhere else. This is because the wormhole manifold \(\mathcal{M}\) is piecewise Schwarzschild with the condition \(a > 2M\) applied to prevent the formation of an event horizon.  The stress-energy tensor being proportional to the delta function makes this situation ideal for the application of the Israel junction conditions~\cite{Israel:1966rt,Blau:1986cw}, also known as the boundary layer formalism. The stress-energy localized at the throat acts as a domain wall, separating the two universes.

On converting to Gaussian normal coordinates for the static case, the metric for the complete space is:
\begin{equation}
ds^2 = -\left(1 - \frac{2M_{\text{WH}}}{R(\ell)}\right) dt^2 + d\ell^2 + R(\ell)^2 \left(d\theta^2 + \sin^2\theta \, d\phi^2\right),
\label{eq:static_metric}
\end{equation}
where the Gaussian normal coordinate \(\ell\) is the proper radius and \(R(\ell)\) satisfies
\begin{equation}
\frac{R}{\ell} = \pm \sqrt{1 - \frac{2M_{\text{WH}}}{R}}.
\label{eq:R_l_relation}
\end{equation}

The wormhole consists of three distinct regions: the outer Schwarzschild spacetimes (denoted as Universe~1 and Universe~2), where the stress--energy tensor is known, and the boundary layer at the throat, where the surface stress--energy must be prescribed either by assumption or guided by observation. The Gaussian normal coordinate \(\ell\) is defined such that \(\ell > 0\) in Universe~1, \(\ell < 0\) in Universe~2, and \(\ell = 0\) exactly at the throat~\cite{Dent:2020nfa}.

To visualize the wormhole, we embed a spatial slice (\(t = \text{const}\), \(\theta = \pi/2\)) into Euclidean space \((r, \phi, z)\). The induced 2D line element is:
\begin{equation}
ds^2 = \left(1 - \frac{2M_\text{WH}}{r} \right)^{-1} dr^2 + r^2 d\phi^2,
\label{eq:dsSquared}
\end{equation}
which can be embedded in \(\mathbb{E}^3\) by identifying a function \(z(r)\) satisfying:
\begin{equation}
\left( \frac{dz}{dr} \right)^2 = \frac{1}{1 - \frac{2M_\text{WH}}{r}} - 1 = \frac{2M_\text{WH}}{r - 2M_\text{WH}},
\label{eq:dzdr}
\end{equation}
yielding the familiar ``flaring throat'' shape~\cite{Visser:1989kh}, as in Figure~\ref{fig:embedding}.

To evaluate the physical properties of the wormhole throat, we model it as a timelike hypersurface across which the spacetime is joined discontinuously. The stress–energy tensor is written as:
\begin{equation}
T^{\mu\nu}(x) = S^{\mu\nu}(x)\,\delta(\ell) + T^{\mu\nu}_+\,\Theta(\ell) + T^{\mu\nu}_-\,\Theta(-\ell),
\label{eq:TmunuGeneral}
\end{equation}
where \(S^{\mu\nu}\) encodes the surface stress--energy on the shell, \(\delta(\ell)\) localizes the curvature at the shell, and \(\Theta(\ell)\) is the Heaviside step function with \(\Theta(0)=0\). In the context of Einstein gravity, the matter required to support a wormhole throat necessarily violates at least one of the classical energy conditions, depending on the chosen equation of state~\cite{Hawking:1973uf}. This can lead to instabilities in some cases~\cite{Buniy:2005vh,Buniy:2006xf}.

We adopt the spacetime signature $(-,+,+,+)$ and use surface \emph{tension} $\vartheta$ (so that $\vartheta=-p$, where $p$ is the tangential surface pressure). For a static, spherically symmetric, and reflection–symmetric throat the momentum flux normal to the shell vanishes, $T^{\ell\mu}=0$, and Israel’s junction conditions yield a diagonal surface stress–energy
\begin{equation}
S^{i}{}_{j}=\mathrm{diag}(-\sigma,-\vartheta,-\vartheta),
\end{equation}
with
\begin{equation}
\sigma = -\frac{1}{2\pi a}\sqrt{1-\frac{2M_{\rm WH}}{a}},\qquad
\vartheta = -\,\frac{1}{4\pi a}\,\frac{1-\frac{M_{\rm WH}}{a}}{\sqrt{1-\frac{2M_{\rm WH}}{a}}}\,.
\end{equation}
(Equivalently, $S^{i}{}_{j}=\mathrm{diag}(-\sigma,p,p)$ with $p=-\vartheta$.)
where \(\sigma\) is the surface energy density.

\section{Radial Infall and Wave Emission}

Dent, et al. \textit{et al.}~\cite{Dent:2020nfa} analyzed the gravitational wave (GW) signatures generated by a black hole (BH) interacting with a traversable wormhole (WH) constructed using Visser's spacetime surgery method~\cite{Visser:1989kg}. Since the WH spacetime is identical to Schwarzschild for all $r>a$, with $a>2M_{\text{WH}}$ to avoid horizons, the initial inspiral signal in Universe~1 is indistinguishable from a standard BH--BH merger until the infalling BH reaches the throat. In Schwarzschild spacetimes, the photon sphere at $r=3M$ plays a key role in shaping the late-inspiral and ringdown waveform, which explains why the signal up to that stage remains unchanged. The interaction model assumes: (i) the throat consists of exotic matter with negative surface energy density $\sigma<0$, (ii) this exotic matter is confined to the shell, and (iii) the only coupling to the infalling BH is gravitational. Because $\sigma<0$, the effective gravitational interaction between the positive-mass BH and the shell matter is repulsive, as follows from the Israel junction conditions \cite{Visser:1989kh,Visser:1989kg}, so we assume the throat matter is momentarily displaced  and then relaxes back with minimal disturbance as the BH transits the throat.

Following the transit, the BH emerges into Universe~2, either escaping to infinity or entering a bound, radial orbit involving multiple throat crossings (discussed later). The associated GW signal consists of a pulse as the BH approaches the throat, followed by a sudden fade-out as it enters the WH. Its emergence into Universe~2 results in a pulse on that side of the throat. If the BH undergoes further throat crossings, this pattern repeats. These features yield distinctive GW signatures such as spectral gaps, repeated pulse cycles, and abrupt transitions, which may be probed by matched filtering against existing LIGO data.

Dent \textit{et al.} focused on bound orbits with angular momentum, including zoom-whirl--like dynamics, and assume symmetric visibility of both WH mouths. In contrast, the present study considers a radial infall of a stellar-mass Schwarzschild BH into a non-spinning traversable wormhole. Since our wormhole is structurally identical to the one considered in their work, all of their assumptions about the wormhole and the nature of the interaction between the black hole and the exotic matter are applicable in our case.

We consider a representative example in which the black hole has mass $M_{\text{BH}} = 5M_\odot$ and the effective mass of the wormhole, on both sides of the throat, is $M_{\text{WH}} = 200M_\odot$, where $M_\odot$ is the mass of the Sun. The wormhole throat is located at $a = 3M_{\text{WH}}$, and yields a cross-sectional area approximately 3600 times larger than the event horizon of the infalling black hole. Under this large area ratio, the passage of the black hole through the throat can be reasonably modeled as a small perturbation to the wormhole geometry, and the traversing BH is not expected to accrete exotic matter, as it is repelled by the positive mass BH. Future studies could explore the effects on motion arising from different wormhole--black hole mass ratios.
 
According to the work of Diemer and Smolarek~\cite{Diemer:2013hgn}, there are four types of orbits that can result from the interaction between a test particle and a wormhole. A Two World Exit (TWE) orbit occurs when the particle enters the wormhole and exits to infinity into another part of the same universe or an entirely different one. A Two World Bound (TWB) orbit describes a gravitationally bound trajectory in which the particle oscillates between the two mouths of the wormhole. A Bound Orbit (BO) is confined near the throat but never crosses to the other side, while an Escape Orbit (EO) occurs when the particle comes from radial infinity, swings around the wormhole, and returns to radial infinity within the same universe.

\begin{figure}[h]
    \centering
    \includegraphics[width=0.8\linewidth]{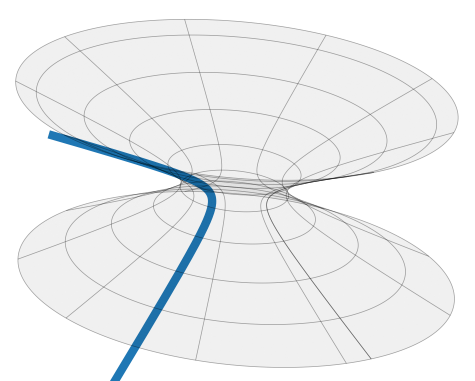}
    \caption{Two‐World Bound (TWB) orbit for radial infall: the black hole begins at radial infinity and plunges along the $z$‐axis with zero angular momentum, traverses the wormhole throat, and thereafter remains bound in a straight‐line, oscillatory trajectory through the throat, indefinitely shuttling between the two universes without energy loss.}
    \label{fig:embed_orbit}
\end{figure}

In this study, we examine a case corresponding to a TWB orbit in the presence of a wormhole as the central object as depicted in Figure \ref{fig:embed_orbit}. We assume idealized conditions in which the black hole does not lose energy through gravitational radiation. Under these assumptions, the orbit remains stable, and the black hole continues to oscillate indefinitely between the two spacetime regions, repeatedly transiting the wormhole throat without settling.

Consider the trajectory of the black hole (BH) starting from rest at a finite distance \( r_i \) and falling radially toward the center of the wormhole (WH). The initial distance \( r_i \) is related to the conserved energy \( E \) by  Equation (3.97) in Chandrasekhar's book~\cite{Chandrasekhar:1985kt}:
\begin{equation}
\left( \frac{dr}{d\tau} \right)^2 = \frac{2M_{WH}}{r} - (1 - E^2)
\label{eq:drdtau}
\end{equation}
where \( \tau \) is the proper time, \( r \) is the radial Schwarzschild coordinate, \( r = r_i \) when \( \dot{r} = 0 \), and \( E \) is the conserved energy per unit mass given by:

\begin{equation}
\left(1 - \frac{2M_{WH}}{r} \right) \frac{dt}{d\tau} = E
\label{eq:energy}
\end{equation}

The equations of motion are most conveniently integrated using a new variable \( \eta \), defined by:

\begin{equation}
r = \frac{M_{WH}}{1 - E^2}(1 + \cos \eta) = \frac{2M_{WH}}{1 - E^2} \cos^2\left(\frac{\eta}{2}\right) = r_i \cos^2\left(\frac{\eta}{2}\right)
\label{eq:requals}
\end{equation}
Note that \( \eta = 0 \) corresponds to \( r = r_i \) and \( \eta = \pi \) corresponds to \( r = 0 \), where the wormhole would reach the singularity (though in our model, the singularity is omitted due to the presence of the wormhole throat).

In terms of \( \eta \), the equation for \( \frac{dt}{d\tau} \) becomes:

\begin{equation}
\frac{dt}{d\tau} = \frac{E \cos^2\left(\frac{\eta}{2}\right)}{\cos^2\left(\frac{\eta}{2}\right) - \cos^2\left(\frac{\eta_H}{2}\right)}
\label{eq:dtdtau}
\end{equation}
where \( \eta_H \) is the value of \( \eta \) at which \( r = 2M_{WH} \), i.e., where the event horizon would have been located given by \( \eta_H = 2 \sin^{-1} E \).

Using Equation (2.8.2) from Weinberg's \textit{Gravitation and Cosmology}~\cite{Weinberg:1972kfs}, the stress-energy tensor for the black hole treated as a point particle of mass \( M_{BH} \) is given by:

\begin{equation}
T^{\mu \nu}(\vec{x}, t) = \frac{M_{BH}}{\sqrt{-g}} \frac{dx^\mu}{dt} \frac{dx^\nu}{dt} \frac{dt}{d\tau} \, \delta^3(\vec{x} - \vec{x}_{BH}(t))
\label{eq:TmunuDef}
\end{equation}

Since the wormhole spacetime has vacuum energy outside the throat and we assume negligible interaction between the BH and the exotic matter at the throat, this expression provides the sole source of stress energy. The \( T^{00} \) component can be evaluated by assuming motion along the equatorial plane (\( \theta = \pi/2 \)), so that \( \sqrt{-g} = r^2(\eta) \), giving:

\begin{align}
T^{00}(\eta, x, y, z) &= \frac{M_{BH}}{\sqrt{-g}} \frac{dx^0}{dt} \frac{dx^0}{dt} \frac{dt}{d\tau} \, \delta^3(\vec{x} - \vec{x}_{BH}(t)) \nonumber \\
&= \frac{M_{BH} \, \frac{dt}{d\tau}}{\sqrt{-g}} \delta^3(\vec{x} - \vec{x}_{BH}(t)) \nonumber \\
&= \frac{E (1 - E^2)^2 M_{BH} \, \sec^2\left(\frac{\eta}{2}\right)}{2 M_{WH}^2 \left(-1 + 2E^2 + \cos\eta\right)} \delta^3(\vec{x} - \vec{x}_{BH}(t))
\label{eq:Tzerozero}
\end{align}

\begin{figure}[H]
    \centering
    \includegraphics[width=1\linewidth]{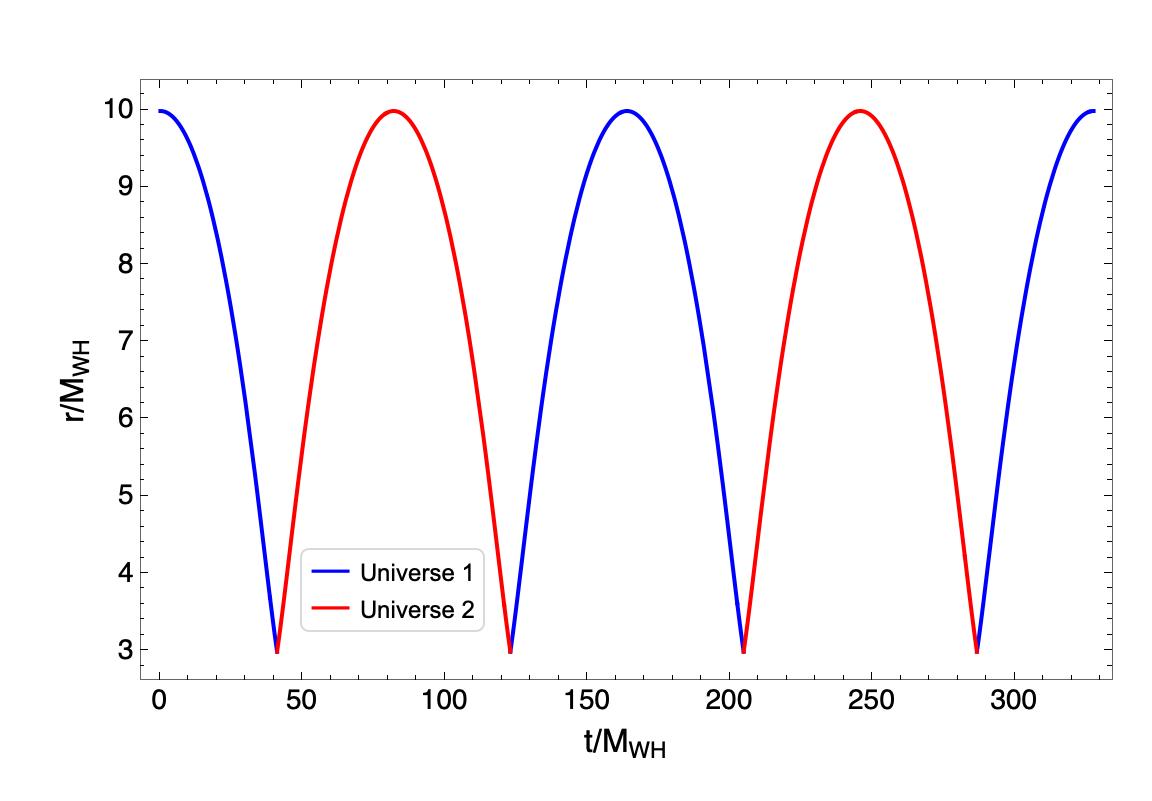}
    \caption{Variation of \( r \) with \( \frac{t}{M_{\mathrm{WH}}} \). The black hole starts at \( r = r_i \) in Universe~1 and falls radially into the wormhole. It reaches the throat at \( t = t_{\mathrm{throat}} \approx 8.19 \times 10^{3} \) in geometrized units, corresponding to approximately \( 0.05\,\mathrm{s} \) for \( M_{\mathrm{WH}} = 200\,M_\odot \). It then traverses to Universe~2, completes a full oscillation, and returns to Universe~1. The cycle continues indefinitely in the absence of dissipative effects.}
    \label{fig:trajectory}
\end{figure}

The trajectory of the black hole through the wormhole is shown in Figure~\ref{fig:trajectory}. We plot the radial Schwarzschild coordinate as a function of \( t \) as the black hole transitions from Universe 1 to Universe 2, and back again.

Table~\ref{tab:valuesUsed} summarizes the numerical values adopted for each variable in our calculations. We use parameters inspired by GW150914, the first direct detection of gravitational waves produced by the merger of two black holes. However, instead of the original component masses (approximately $36\,M_\odot$ and $29\,M_\odot$), we take representative values of $M_{\rm WH} = 200\,M_\odot$ and $M_{\rm BH} = 5\,M_\odot$. We further assume that the wormhole--black hole interaction occurs at a luminosity distance of $D_L = 500~\mathrm{Mpc}$, corresponding to the estimated distance of GW150914. Finally, we take $E = 0.949$, as this value is known to produce a TWB orbit~\cite{Diemer:2013hgn}.

We now proceed to calculate the gravitational waves radiated from our wormhole--black hole interaction. As outlined in~\cite{Flanagan:2005yc}, gravitational waves are treated as small perturbations propagating on a fixed background spacetime and satisfying a wave equation in the linearized approximation. In the far-field, or radiation zone, these perturbations can be described in the transverse-traceless (TT) gauge, where the physical degrees of freedom appear as the plus \(\left(h_{+}\right)\) and cross \(\left(h_{\times}\right)\) polarizations~\cite{Chakrabarty:1999aa}. We adopt this framework to characterize the gravitational radiation emitted by the black hole as it traverses the wormhole geometry.

To model the gravitational wave emission from our wormhole--black hole system, we adopt the standard mass quadrupole formalism as outlined in~\cite{Maggiore:2007ulw}. At leading order, the gravitational radiation is determined by the second time derivative of the mass quadrupole moment \( M_{ij} \) of the source. In the far-field limit and within the transverse-traceless (TT) gauge, the plus and cross polarizations of the emitted gravitational waves are given by
\begin{align}
h_+(t, \theta, \phi) &= \frac{1}{D} \frac{G}{c^4} \bigg[
(\ddot{M}_{11} - \ddot{M}_{22}) \cos^2 \phi 
+ 2 \ddot{M}_{12} \sin \phi \cos \phi
\bigg] \notag \\
&\quad \cos^2 \theta + \frac{1}{D} \frac{G}{c^4} \bigg[
\ddot{M}_{33} 
- \ddot{M}_{11} \sin^2 \phi 
- \ddot{M}_{22} \cos^2 \phi
\bigg], \notag \\
h_\times(t, \theta, \phi) &= \frac{1}{D} \frac{G}{c^4} \bigg[
(\ddot{M}_{11} - \ddot{M}_{22}) \sin \phi \cos \phi
\bigg] \cos \theta \notag \\
&\quad + \frac{1}{D} \frac{G}{c^4} \bigg[
\ddot{M}_{12} (\cos^2 \phi - \sin^2 \phi)
\bigg] \cos \theta.
\label{eq:hplushcross}
\end{align}
as summarized in Equation (3.72) of~\cite{Maggiore:2007ulw}. Here, $\theta$ and $\phi$ are the spherical coordinate angles and $D$ is the distance to the object (or system of celestial objects) emitting the waves. In our case, the mass quadrupole tensor \( M_{ij} \) is constructed directly from the black hole's trajectory through the wormhole, with its components expressed as \( M^{ij} = \frac{1}{c^2} \int d^3x \, T^{00}(\vec{x}) x^i x^j \). Taking appropriate derivatives with respect to proper time and projecting onto the TT frame, and assuming $\theta=\frac{\pi}{2}$, $\phi=0$ , we compute the gravitational waveforms generated during the black hole’s traversal of the wormhole:

\begin{align}
h_{+}(\eta) &=
\frac{1}{16\,c^{4}(M_{\mathrm{BH}} + M_{\mathrm{WH}})\,D}
\Bigg[
5 G\, M_{\mathrm{BH}}\, M_{\mathrm{WH}}^{2}\notag
\\[-2pt]
&\left( -1 + 2\cos\eta + \cos 2\eta - 2\cos\eta_{H} \right)\notag
\\[-2pt]
&\left( \cos\eta - \cos\eta_{H} \right)
\sec^{6}\!\left( \frac{\eta}{2} \right)
\Bigg],\notag
\\[4pt]
h_{\times}(\eta) &= 0.
\end{align}

As shown in the calculations, the cross polarization $h_\times$ vanishes, leaving only the plus polarization $h_+$  nonzero. Figure 4 presents the plus polarization (as seen in \textit{Universe 1}) plotted against the parameter $\eta$ for two complete cycles of the black hole traversing the wormhole throat. Specifically, this trajectory involves the black hole starting in \textit{Universe 1}, crossing into \textit{Universe 2}, returning to \textit{Universe 1}, crossing back into \textit{Universe 2}, and finally returning to \textit{Universe 1}.

\begin{figure}[h]
    \centering
    \includegraphics[width=1.1\linewidth]{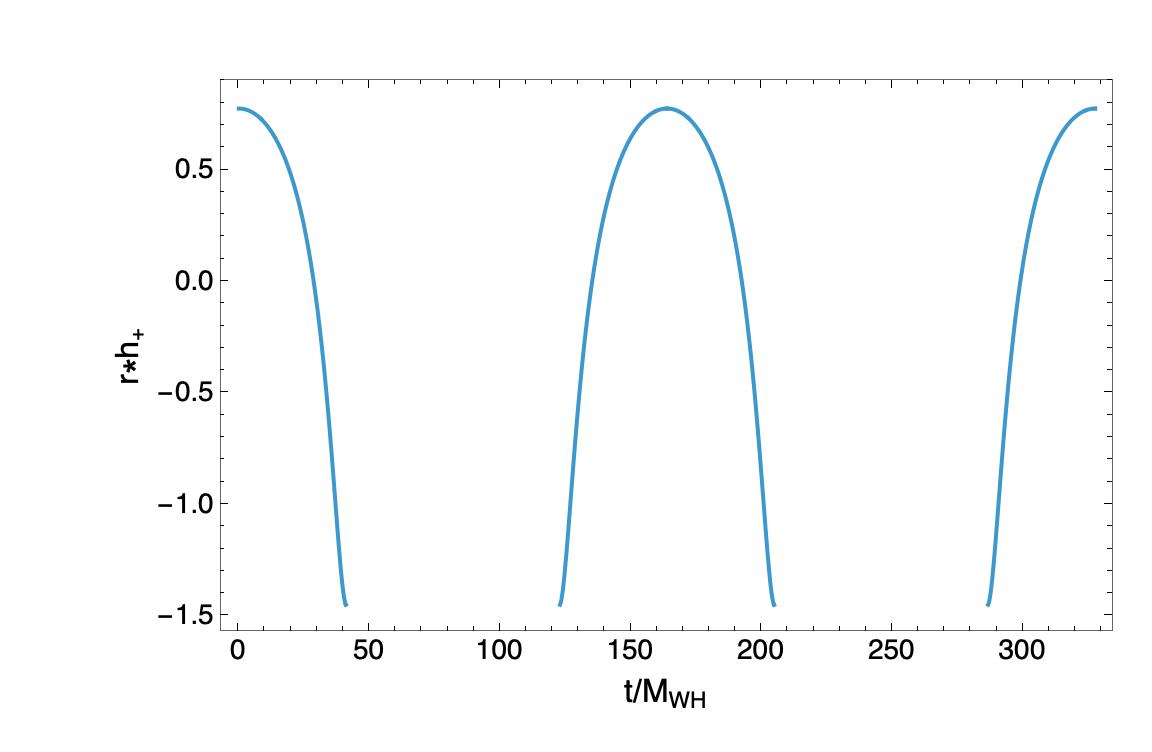}
    \caption{Mass quadrupole radiation \(rh_+\) versus the time \(\frac{t}{M_{WH}}\). The pulse segments, corresponding to the black hole's traversal through Universe 1 while the  gaps indicate the black hole's passage into Universe 2.}
    \label{fig:massQuadRad}
\end{figure}

The resulting waveform exhibits the characteristic \textit{pulse} pattern when the black hole is in \textit{Universe 1}, where the amplitude rapidly increases as the black hole gains kinetic energy while approaching the wormhole throat. In contrast, the \textit{anti-pulse} appears as distinct gaps in the waveform when the black hole is in \textit{Universe 2}. These gaps correspond to regions where the black hole has crossed the throat and is moving away, resulting in a sharp drop in the observed signal amplitude. This alternating structure captures the distinct gravitational signatures produced by the black hole's repeated traversal, encoding valuable information about the throat geometry, mass distribution, and the gravitational interactions within the wormhole.

\begin{table}[H]
\centering
\def\arraystretch{1.5}
\setlength\tabcolsep{0.015\textwidth} 
\begin{tabular}{|c|c|c|c|c|c|}
\hline
\( E \) & \( D \) & \(M_{\rm WH}\) & \( M_{BH} \) & \( \theta_{GW} \) & \( \phi_{GW} \) \\
\hline
0.949 & 500 Mpc & 200 \(M_\odot\) & 5 \(M_\odot\) & \( \frac{\pi}{2} \) & \( 0 \) \\
\hline
\end{tabular}
\caption{Numerical values used. Here, \(D\) is the luminosity distance to the source,
\(M_{\rm WH}\) is the wormhole mass, and \(M_{\rm BH}\) is the black hole mass,
with \(M_\odot\) denoting one solar mass. The angles
\(\theta_{\rm GW}\) and \(\phi_{\rm GW}\) are the spherical polar angles (in radians)
that specify the observer’s line of sight used in the gravitational-wave
projection. The wormhole throat is located at an areal radius
\(a = 3M_{\rm WH}\). The black hole is released from an initial areal radius
\(r_i = 10M_{\rm WH}\).}
\label{tab:valuesUsed}
\end{table}

The Current Quadrupole Moment captures the rotational aspects of mass currents within the system. For a purely radial infall without any rotational motion, this tensor is expected to vanish. In our calculations, each component of the current quadrupole tensor was found to be exactly zero, confirming that there is no current quadrupole radiation in this scenario. This result aligns with the physical expectation that a straight, non-rotating trajectory through the wormhole cannot generate the off-diagonal stress-energy components necessary for current quadrupole radiation.

The \textit{Mass Octupole Moment} represents the next-to-leading order term in the gravitational wave expansion beyond the mass quadrupole. It captures the more complex, higher-order asymmetries in the mass distribution of a system. 

The mass octupole tensor \(M^{ijk}\) is defined as the third-order spatial moment of the energy density,
\begin{equation}
M^{ijk}(t) = \frac{1}{c^2}\int d^3x \, T^{00}(t,\vec{x}) \, x^i x^j x^k,
\label{eq:def_Mijk}
\end{equation}
which is completely symmetric in its indices. For a point particle of mass \(m\) at position \(\vec{x}(t)\), this reduces to \(M^{ijk} = m\,x^i x^j x^k\).

The corresponding radiation term, typically denoted as \(O^{klm}\), is obtained by removing all traces from \(M^{ijk}\) as given by
\begin{equation}
O^{klm} = M^{klm} - \frac{1}{5} \left( \delta^{kl} M^{n}{}_{n}{}^{m} + \delta^{km} M^{n}{}_{n}{}^{l} + \delta^{lm} M^{n}{}_{n}{}^{k} \right),
\label{eq:bigO}
\end{equation}
where the repeated spatial index \(n\) is summed over. This ensures that \(O^{klm}\) is symmetric and traceless. 

The resulting gravitational wave component can then be expressed as in Eqn.~(3.141) of~\cite{Maggiore:2007ulw},
\begin{equation}
(h_{ij}^{TT})_{\text{oct}} = \frac{1}{D} \frac{2G}{3c^5} \Lambda_{ij,kl}(\hat{n}) n_m \ddot{O}^{klm},
\label{eq:hTT}
\end{equation}
where \(\Lambda_{ij,kl}(\hat{n})\) is the transverse, traceless projection operator, ensuring that only radiative degrees of freedom are included. Unlike the mass quadrupole, the mass octupole radiates at higher frequencies and with a smaller amplitude, as the power radiated scales as \(\frac{v^2}{c^2}\) relative to the quadrupole. This means that the mass octupole contribution is typically much weaker, but it can become significant for systems with strong higher-order asymmetries or non-linear interactions.

Solving Eq.~\ref{eq:hTT} for the mass octupole radiation in our setup results in a vanishing \(h_\times\) and yields Eq.~\ref{eq:mass_octupole_waveform_compact} for \(h_+\).

\begingroup
\small
\begin{equation}
\begin{aligned}[t]
h_+(\eta) &=
\frac{1}{
256\, c^5\, (M_{\rm BH} + M_{\rm WH})^2\, \sqrt{ D^{\,3} / M_{\rm WH} }
}
\Bigg[
\\&
5 \sqrt{5}\, G\, M_{\rm BH}\, (M_{\rm BH} - M_{\rm WH})\, M_{\rm WH}^2 \,
(\cos\eta - \cos\eta_H)
\\
&\quad\times
\Big(
5 + 6\cos(2\eta) + \cos(3\eta)
 - 4\cos\eta_H + \cos(2\eta_H)
 \\
&\quad - \cos\eta\,(-4 + 12\cos\eta_H + \cos(2\eta_H))
\Big)
\\
&\quad\times
\sec^8\!\left(\frac{\eta}{2}\right)\,
\tan\!\left(\frac{\eta}{2}\right)
\Bigg].
\end{aligned}
\label{eq:mass_octupole_waveform_compact}
\end{equation}
\endgroup

\begin{figure}[H]
    \centering
    \includegraphics[width=1.1\linewidth]{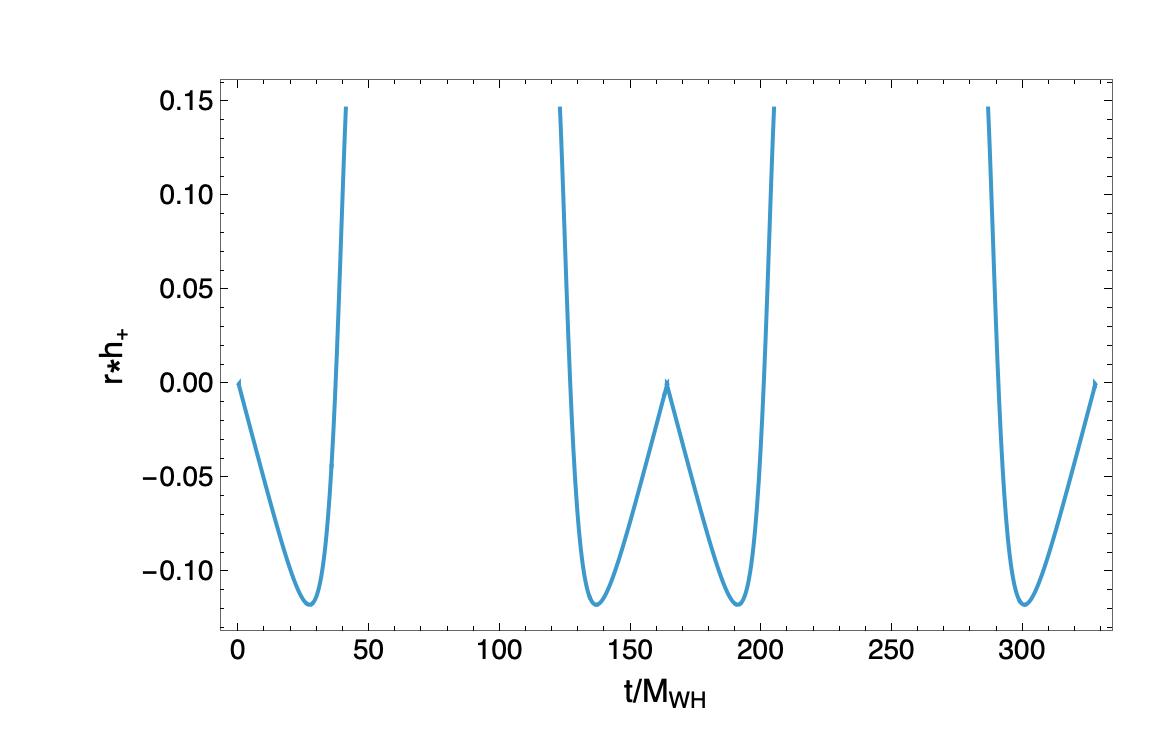}
    \caption{ Mass octupole radiation \(rh_+\) versus the time. The segments correspond to the black hole traversing Universe 1 while the gaps indicate the black hole moving in Universe 2.}
    \label{fig:hplus_oct}
\end{figure}

In Figure~\ref{fig:hplus_oct}, we plot \(h_+\) revealing the repeated traversal of the black hole through the wormhole throat, cycling between Universe 1 and Universe 2. 

\section{Frequency-Domain Analysis and Ground-Based Detector Comparison}
\label{sec:asd}

In this section we describe the conversion of the time-domain waveform $h(t)$
from Sec.~\ref{fig:trajectory} into an amplitude spectral density (ASD), and
compare it with representative ground-based gravitational-wave detector
strain sensitivities. We follow standard Fourier and power-spectral-density
conventions \cite{Flanagan:2005yc,Maggiore:2007ulw}.

\subsection{Preprocessing of $h(t)$}
The strain time series is sampled at uniform time steps in physical seconds.\footnote{
If intermediate calculations are performed in units of $t/\Mwh$, we convert
to SI units using $t_{\rm phys}=(G\Mwh/c^3)\,(t/\Mwh)$.}
During intervals in which the infalling black hole resides in
\emph{Universe~2}, the gravitational radiation is causally disconnected from
an observer in Universe~1. We therefore set
\[
h(t)=0 \quad \text{during all Universe~2 intervals,}
\]
which yields the physically observable strain for a Universe~1 detector.

\subsection{Windowing and discrete Fourier transform}
Let $h_n \equiv h(n\Delta t)$ for $n=0,\dots,N-1$, with total duration
$T=N\Delta t$. Because the waveform is finite in time, a direct Fourier
transform would introduce spectral leakage due to sharp boundaries at the
segment endpoints. To suppress these artificial high-frequency artifacts,
we apply a Hann window $w_n$ prior to the transform.

The Hann window smoothly tapers the signal to zero at the boundaries while
preserving broadband spectral content. To compensate for the associated
reduction in signal power, we normalize by
\[
U \equiv \frac{1}{N}\sum_{n=0}^{N-1} w_n^2 .
\]
We then compute the discrete Fourier transform (implemented using a real FFT)
with the continuous-time scaling
\begin{equation}
\label{eq:rfft}
\tilde h_k \;=\; \frac{\Delta t}{\sqrt{U}} \sum_{n=0}^{N-1}
\bigl(w_n h_n\bigr)\, e^{-2\pi i kn/N}\,,
\qquad k=0,\dots,\tfrac{N}{2}.
\end{equation}
The corresponding positive frequency bins are
\[
f_k \;=\; \frac{k}{N\,\Delta t}\,.
\]

\subsection{Single-sided PSD and ASD}
The one-sided strain power spectral density (PSD) is defined as
\begin{equation}
\label{eq:psd}
S_h(f_k) \;=\;
\begin{cases}
\displaystyle \frac{2}{T}\,|\tilde h_k|^2, & 1 \le k \le \tfrac{N}{2}-1,\\[6pt]
\displaystyle \frac{1}{T}\,|\tilde h_k|^2, & k=0 \text{ or } k=\tfrac{N}{2},
\end{cases}
\end{equation}
and the corresponding amplitude spectral density is
\begin{equation}
\label{eq:asd}
{\rm ASD}(f_k) \;=\; \sqrt{S_h(f_k)} \quad [{\rm Hz}^{-1/2}]\, .
\end{equation}
The factor of $2$ accounts for folding negative-frequency power into the
positive-frequency band, while the window normalization ensures an unbiased
estimate of spectral power.

\subsection{Comparison with ground-based detectors}
For context, we overlay the resulting ASD with representative design
sensitivity curves for current and future ground-based detectors, including
Advanced LIGO, the Einstein Telescope (ET-D), and Cosmic Explorer (CE1 and
CE2). Since the present system produces $h_\times=0$ by symmetry, the detector
response reduces to
\[
h_{\rm det}(t)=F_+(\hat n,\psi)\,h_+(t),
\]
and we set $F_+=1$ for an optimally oriented source.

The comparison is intended to provide an order-of-magnitude reference rather
than a detectability claim. For the parameters considered here, the wormhole
signal remains well within the design sensitivity of existing and planned
ground-based detectors across the relevant frequency band.

\begin{figure*}[t]
    \centering
    \includegraphics[width=\textwidth]{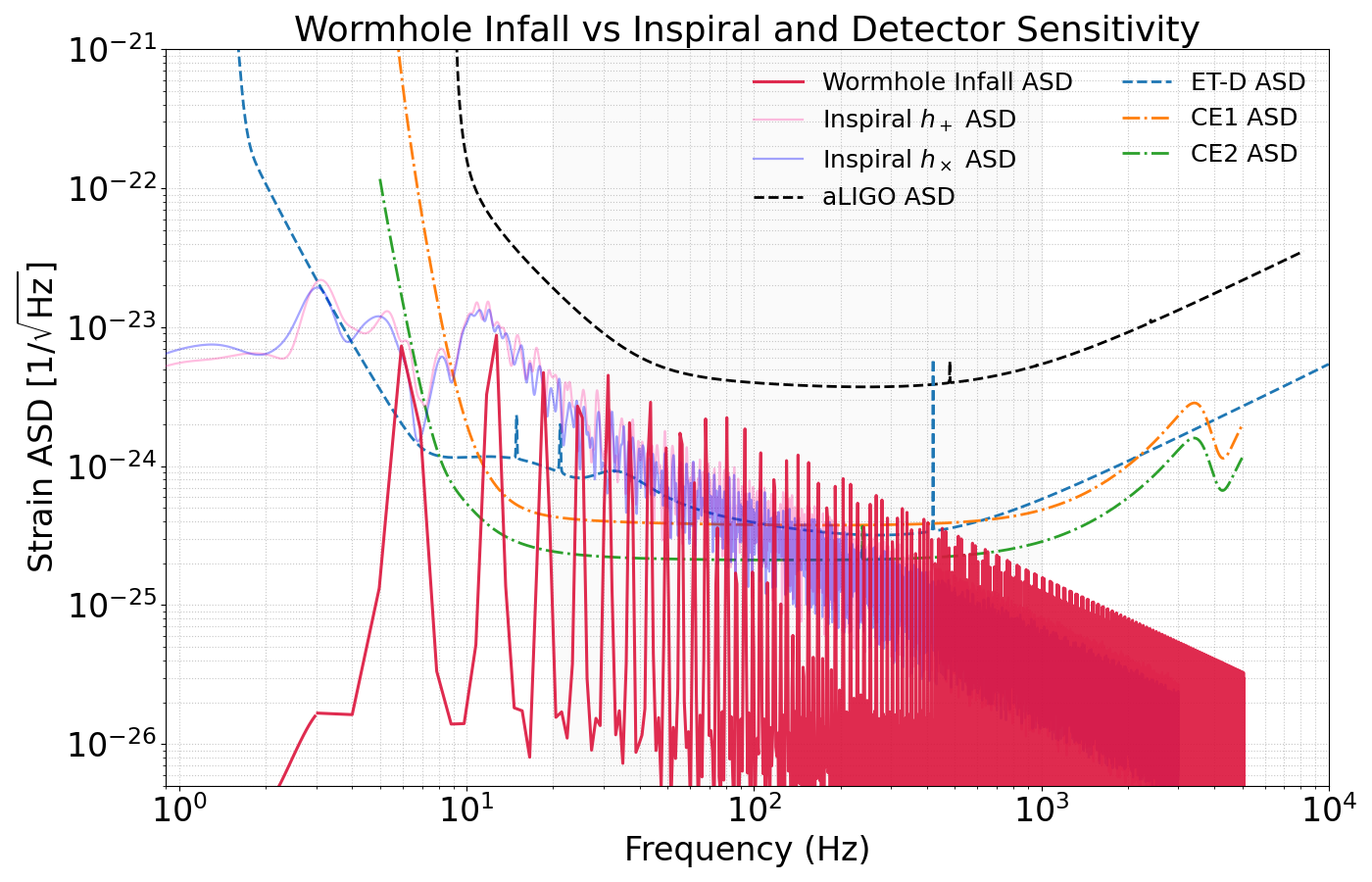}
    \caption{Single-sided amplitude spectral density (ASD) of the wormhole infall waveform (red), compared with the inspiral contribution from a black hole–wormhole system (light and dark purple, corresponding to the plus and cross polarizations), and representative ground-based detector sensitivities (dashed). The inspiral signal is computed for a system consisting of a $5M_\odot$ black hole orbiting a $200M_\odot$ traversable wormhole (mass ratio $\sim 40{:}1$) at a luminosity distance of $500\mathrm{Mpc}$ and inclination angles $\theta=\phi=\pi/4$. The inspiral exhibits a smooth, broadband spectrum characteristic of quasi-circular orbital motion, while the wormhole infall signal displays pronounced oscillatory structure due to repeated passages through the wormhole throat, producing burst-like features and spectral modulation. A Hann window is applied to the full time series, and intervals corresponding to emission in the second asymptotic region (Universe 2) are set to zero.} \label{fig:asd_overlay}
\end{figure*}

\section{Summary and Discussion}

Next-generation gravitational-wave observatories are expected to probe compact
object mergers from deep cosmic volumes, enabling increasingly precise tests
of strong-field gravity. In particular, such detectors offer a promising avenue
for exploring exotic compact objects and nontrivial spacetime geometries,
including traversable wormholes, which may provide insight into the quantum
structure of gravity and the limits of classical general relativity
\cite{Kalogera:2021bya}.

In this work, we investigated the gravitational-wave signatures produced by the
radial infall of a stellar-mass black hole into a thin-shell Schwarzschild
wormhole constructed via spacetime surgery. By modeling the black hole as a
test particle and neglecting backreaction on the wormhole throat, we obtained
analytic expressions for the trajectory and derived the associated gravitational
radiation using a multipolar expansion. Our analysis included the mass
quadrupole, mass octupole, and current quadrupole moments, and demonstrated that
symmetry considerations suppress the cross polarization and all current-type
contributions. The resulting waveform exhibits a characteristic pulse  as the black hole approaches, traverses, and recedes from the
wormhole throat. This behavior is qualitatively distinct from the inspiral and
merger signals of conventional black-hole binaries and provides a potential
discriminant for wormhole-induced gravitational-wave events.

A key result of this study is the frequency-domain characterization of the
signal. We computed the single-sided amplitude spectral density (ASD) of the
wormhole-induced quadrupole waveform and compared it directly with
representative ground-based detector strain sensitivities. Our analysis shows
that, for the system parameters considered here, the ASD of the radial infall
signal can intersect the sensitivity bands of current and next-generation
ground-based detectors for source distances up to $\lesssim 500\,\mathrm{Mpc}$,
assuming an optimally oriented source. This suggests that radial black
hole–wormhole infall events within this distance range may be detectable by
ground-based observatories, motivating further searches for nonstandard burst-like signals in gravitational-wave data.

While the present analysis focuses on strictly radial infall, black-hole
trajectories in realistic astrophysical environments need not be purely radial
or circular. The Kozai–Lidov mechanism \cite{Kozai:1962zz,Lidov:1962wjn}, for example,
predicts that perturbations from a third massive body can drive large
eccentricities and inclination oscillations. In a black hole–wormhole system,
such effects could lead to highly eccentric encounters or repeated throat
crossings, producing more complex and potentially higher-amplitude gravitational
wave bursts. These scenarios introduce the possibility of chaotic dynamics and
extreme mass-ratio interactions that extend beyond the idealized configurations
considered here.

More broadly, wormholes provide a unique testing ground for fundamental physics,
challenging standard notions of causality, topology, and energy conditions.
Gravitational-wave observations offer a direct, dynamical probe of these exotic
spacetimes and may ultimately provide empirical evidence for physics beyond
classical general relativity \cite{Hayward:1998pp,Farrah:2023opk}. In addition, quantum effects
associated with wormholes have been suggested as a possible ingredient in
addressing the cosmological constant problem \cite{Klebanov:1988eh}. Although the
thin-shell wormholes studied here require exotic matter for stability, recent
work in modified and quadratic gravity theories has demonstrated the existence
of traversable wormhole solutions that evade classical energy-condition
violations \cite{Duplessis:2015xva,Dent:2016efw}.

This work represents an initial step toward a systematic exploration of
gravitational-wave signatures from traversable wormholes. Future extensions
could include rotating (Kerr-like) wormhole geometries, nonradial and eccentric
trajectories, and the inclusion of higher-order multipoles and relativistic
backreaction effects. A more comprehensive survey of waveform morphologies and
their detectability across detector networks will be essential for assessing
the astrophysical relevance of black hole–wormhole interactions and for guiding
targeted searches for exotic compact objects in current and future
gravitational-wave data.

{\bf Acknowledgements --} JBD acknowledges support from the National Science Foundation under grant no. PHY2412995. JBD thanks the Mitchell Institute at Texas A\&M University for its hospitality
where part of this work was completed.

\bibliography{main}

\end{document}